\documentstyle[aps,preprint,tighten,epsfig]{revtex}
\def\beq{\begin{equation}}
\def\eeq{\end{equation}}
\def\beqa{\begin{eqnarray}}
\def\eeqa{\end{eqnarray}}
\def\MeV{\nobreak\,\mbox{MeV}}
\def\GeV{\nobreak\,\mbox{GeV}}

\def\pli{p^\prime}
\def\mli{{M^\prime}^2}
\begin{document}
\title{\sc  $D$ and $\rho$ Mesons: Who resolves whom?}
\author {M.E. Bracco$^1$, M. Chiapparini$^1$, A. Loz\'ea$^1$, F.S. 
Navarra$^2$ and M. Nielsen$^2$\\
\vspace{0.3cm}
{\it $^1$Instituto de F\'{\i}sica,  Universidade do Estado do Rio de Janeiro,}
 \\ 
{\it Rua S\~ao Francisco Xavier 524, Maracan\~a,  20559-900, Rio de Janeiro, 
RJ, Brazil}\\
\vspace{0.3cm}
{\it $^2$Instituto de F\'{\i}sica, Universidade de S\~{a}o Paulo, } \\
{\it C.P. 66318,  05389-970 S\~{a}o Paulo, SP, Brazil}}
\maketitle
\vspace{1cm}

\begin{abstract}
The $DD\rho$ form factor is evaluated in a QCD sum rule calculation
for both $D$ and $\rho$ off-shell mesons. 
We study the double Borel sum rule for the three
point function of two pseudoscalar and one vector meson currents.
We find that the momentum dependence of the form factors is very different
if the $D$ or the $\rho$ meson is off-shell, but they lead to the same 
coupling constant in the $DD\rho$ vertex. We discuss two different
approaches to extract the $DD\rho$ coupling constant.
\\
PACS numbers 14.40.Lb,~~14.40.Nd,~~12.38.Lg,~~11.55.Hx
\\

\end{abstract}

\vspace{1cm}

Among the basic things that we want to know about hadrons are their sizes, 
or, equivalently, their form factors \cite{bad,pauli}. 
The size of a hadron depends on how
we ``look'' at it. The most extensively studied particle is the nucleon, 
which has been probed mainly by photons. In lower energy experiments, where 
also the four momentum transfer ($q^2$) is low,  it was possible to determine 
the electromagnetic form factor  (and the charge radius) of the nucleon. In 
higher energies experiments and very large values of $q^2$ a very different 
picture of the nucleon emerged, in which it is made of pointlike particles, 
the quarks. From these  observations one may conclude that, when probing the 
nucleon, nearly on-shell photons ($q^2 \simeq 0$) recognize  sizes whereas 
highly off-shell photons ($q^2 << 0$) do not.  This statement is supported 
by the phenomenologically very successful vector meson dominance hypothesis, 
according to which real photons are with a large probability converted to 
vector mesons (which are extended objects) and then interact with the nucleon. 
Recent HERA data on electron-proton reactions can be well understood 
introducing a ``transverse radius of the photon'', parametrized as \cite{cart} 
\begin{equation}
r_{\gamma} \simeq \frac{1}{\sqrt{Q^2 + m^2}}
\end{equation}
where $Q^2 = - q^2$ and $m$ is the mass of the vector meson considered. This 
empirical formula tells us that for $Q^2 \rightarrow \infty$ the photon is 
pointlike  and ``resolves'' the nucleon target, i.e., identifies its 
pointlike constituents and does not ``see''  the size of the nucleon.
Moreover this formula indicates that for $Q^2 \simeq 0$ and for light mesons
(like the $\rho^0$) the photon has appreciable transverse radius  and therefore 
also identifies the global nucleon extension. Finally, in the above formula we 
may have a heavy vector meson ($J/\psi$ or $\Upsilon$) which will, either 
real or virtual, resolve the nucleon into pointlike constituents. This feature nicely explains why the $J/\psi$ photoproduction cross section grows 
pronouncedly with the (photon-proton) energy whereas the $\rho$ cross section
grows very slowly \cite{cart} . In the former case the compact $J/\psi$ interacts with 
the small $x$ gluons in the protons, which have a fastly growing population. 
In the latter case, the $\rho$ identifies the global and slowly growing 
geometrical size of the proton. 

Along this line of reasoning one might ask i)
how is the nucleon form factor when 
probed by heavy mesons ($J/\psi$, $D$, $D^*$, ...) ?
 and  also ii)  which 
are the form factors of these heavy mesons when probed by light particles such 
as photons, pions and $\rho$ mesons?  
Apart from their intrinsic value as 
fundamental knowledge about nature, the answers to these questions have 
immediate applications to the physics of  quark gluon plasma (QGP). Indeed, 
form factors and coupling constants of all vertices of the type $B B' M_c$ and
$M M' M_c$, where $B(B')$, $M(M')$ and $M_c$ are respectively baryons, light 
mesons and charmed mesons, are relevant for the calculation of  
interaction cross sections of charmed mesons in nuclear matter  
\cite{linko,lkz,ldk,haglin,haga,stt,osl,nnr,lkl}.  

A couple of years ago we started our program of computing the above mentioned 
quantities in the framework of QCD sum rules (QCDSR). Concerning question i) , 
in \cite{nn} we calculated the coupling constant in the vertex 
$N D \Lambda_c$ with the $D$ meson off-shell. In \cite{dnn} we extended the 
calculation performed in \cite{nn} and computed the $Q^2$ dependent form 
factors of the $N D \Lambda_c$ and $N D^* \Lambda_c$ vertices. We have also 
studied the  $D^* D \pi$ and $B^* B \pi$ vertices \cite{ddpi}. One of the 
conclusions of these works 
is that when the off-shell particle in the vertex is heavy, the form 
factor tends to be broader (or harder) as a function of $Q^2$ , which 
means larger cut-off parameters and smaller associated sizes.

In the present work we will further investigate  form factors involving 
heavy mesons in order to extend our previous conclusions. We also want to 
better estimate the uncertainties in the procedure of determining coupling 
constants  with our techniques. In particular, we want to check whether the 
coupling is the same regardless of which particle is off-shell. For these 
purposes we consider the vertex $D D \rho$ and compute form factors and 
coupling constants for the cases where the $D$ is off-shell, the $\rho$ is 
off-shell and then compare the results. Following the QCDSR formalism  
described in our previous works we write the three-point function associated 
with a $DD\rho$ vertex, which is given by
\begin{equation}
\Gamma_\mu^{(D)}(p,\pli)=\int d^4x \, d^4y \, \langle 0|T\{j_+(x)
j_0(y)j_\mu^\dagger(0)\}|0\rangle  
\, e^{i\pli.x} \, e^{-i(\pli-p).y}\; , 
\label{cord}
\end{equation}
for a $D$ off-shell meson, and by
\begin{equation}
\Gamma_\mu^{(\rho)}(p,\pli)=\int d^4x \, d^4y \, \langle 0|T\{j_+(x)
j_\mu^\dagger(y)j_0(0)\}|0\rangle  
\, e^{i\pli.x} \, e^{-i(\pli-p).y}\; , 
\label{corro}
\end{equation}
for a $\rho$ off-shell meson,
where $j_+=i\bar{d}\gamma_5 c$, $j_0=i\bar{c} \gamma_5u $ and
$ j_\mu^\dagger=\bar{u}\gamma_\mu d$ are the interpolating fields for $D^+$, 
$D^0$ and $\rho^+$ respectively with $u$, $d$ and $c$ being the up, down, and 
charm quark fields.

The general expression for the vertex function in Eqs.~(\ref{cord}) and
(\ref{corro}) has two independent structures.
Writing $\Gamma_\mu$ in terms of the invariant
amplitudes associated with these two structures:
\beq
\Gamma_\mu(p,\pli)=\Gamma_1(p^2,{\pli}^2,q^2)p_\mu + \Gamma_2(p^2,{\pli}^2,q^2)
\pli_\mu\;,
\eeq
we can write a double dispersion relation for each one of the invariant
amplitudes $\Gamma_i\,(i=1,2)$, over the virtualities $p^2$ and ${\pli}^2$
holding $Q^2=-q^2$ fixed:
\beq
\Gamma_i(p^2,{\pli}^2,Q^2)=-{1\over4\pi^2}\int_{s_{min}}^\infty ds
\int_{m_c^2}^\infty du {\rho_i(s,u,Q^2)\over(s-p^2)(u-{\pli}^2)}\;,
\label{dis}
\eeq
where $\rho_i(s,u,Q^2)$ equals the double discontinuity of the amplitude
$\Gamma_i(p^2,{\pli}^2,Q^2)$ on the cuts $s_{min}\leq s\leq\infty$,
$m_c^2\leq u\leq\infty$, and where  $s_{min}=m_c^2$ in the case of the $D$ 
off-shell, where the dispersion relation is written in terms of the
two $D$ mesons' momenta, and  $s_{min}=0$ in the case of $\rho$ off-shell,
the dispersion relation being now written in terms of the $\rho$ and the
$D$ meson momenta. 

The double discontinuity of the perturbative contribution reads:
\beq
\rho_1^{pert(D)}(s,u,Q^2)={6(s+Q^2+u)[m_c^2(m_c^2+s+Q^2-u)-Q^2u]\over
[(s+u+Q^2)^2-4su]^{3/2}}\; ,
\label{d1}
\eeq
\beq
\rho_2^{pert(D)}(s,u,Q^2)={12s[m_c^2(m_c^2+s+Q^2-u)-Q^2u]\over
[(s+u+Q^2)^2-4su]^{3/2}}\; ,
\label{d2}
\eeq
for a $D$ off-shell meson, with the integration limit condition:
\beq
(m_c^2+Q^2)(u-m_c^2)\geq sm_c^2\; ,
\eeq
and
\beq
\rho_1^{pert(\rho)}(s,u,Q^2)={3(s+Q^2-u)[m_c^2(m_c^2-s-Q^2-u)+su]\over
[(s+u+Q^2)^2-4su]^{3/2}}\; ,
\label{ro1}
\eeq
\beq
\rho_2^{pert(\rho)}(s,u,Q^2)={-3(s-Q^2-u)[m_c^2(m_c^2-s-Q^2-u)+su]\over
[(s+u+Q^2)^2-4su]^{3/2}}\; ,\label{ro2}
\eeq
for a $\rho$ off-shell meson, with the integration limit condition:
\beq
(s-m_c^2)(u-m_c^2)\geq Q^2m_c^2\; .\label{lro}
\eeq

The phenomenological side of the vertex function is
obtained by considering the contribution of the $\rho$ and one $D$ meson, 
or the two $D$ mesons  states  to the matrix element
in Eqs.~(\ref{cord}) and (\ref{corro}) respectively:
\beqa
\Gamma_\mu^{phen(D)}(p,\pli)&=&-{m_D^2f_D\over m_c}{m_\rho^2\over g_\rho}
g_D(Q^2){1\over p^2-m_\rho^2}{1\over{\pli}^2-m_D^2}\times
\nonumber \\*[7.2pt]
&&\left(-2\pli_\mu+{m_D^2+m_\rho^2+Q^2\over m_\rho^2}p_\mu\right)
+ \mbox{higher resonances}\; ,
\label{phend}
\eeqa
\beqa
\Gamma_\mu^{phen(\rho)}(p,\pli)&=&-{m_D^4f_D^2\over m_c^2}
g_\rho(Q^2){1\over p^2-m_D^2}{1\over{\pli}^2-m_D^2}
\left(\pli_\mu+p_\mu\right)
+ \mbox{higher resonances}\; ,
\label{phenro}
\eeqa
where
\beq
g_D(Q^2)={m_D^2f_D\over m_c}{g_{DD\rho}^{(D)}(Q^2)\over Q^2+m_D^2}\;,
\label{gd}
\eeq
and
\beq
g_\rho(Q^2)={m_\rho^2\over g_\rho}{g_{DD\rho}^{(\rho)}(Q^2)\over Q^2+
m_\rho^2}\;, \label{gro}
\eeq
where $m_D$, $m_\rho$, $f_D$ and $g_\rho$ are the masses and decay constants
of the mesons $D$ and $\rho$ respectively, and $g_{DD\rho}^{(M)}(Q^2)$
is the form factor at the $DD\rho$ vertex when the meson $M$ is off-shell.
The contribution of higher resonances in Eqs.~(\ref{phend})
and (\ref{phenro}) will be taken into account as usual in the standard form of 
continuum contribution from the thresholds $s_0$ and $u_0$ \cite{io2}. 
Finally we perform a double Borel transformation \cite{io2} in both variables
$P^2=-p^2\rightarrow M^2$ and ${P^\prime}^2=-{\pli}^2\rightarrow \mli$, for  
each invariant amplitude, and 
identify the two representations described above.

For consistency we use in our analysis the QCDSR expressions for 
the decay constants appearing in Eqs.~(\ref{phend}) and (\ref{phenro})
up to dimension four in lowest order of $\alpha_s$:
\beq
f_D^2={3m_c^2\over 8\pi^2m_D^4}\int_{m_c^2}^{u_0}du 
{(u-m_c^2)^2\over u}e^{(m_D^2-u)/\mli}\,-\, {m_c^3\over m_D^4}
\langle\bar{q}q\rangle e^{(m_D^2-m_c^2)/\mli}\; ,\label{fd}
\eeq
\beq
{m_\rho^2\over g_\rho^2}={M^2\over 8\pi^2}\left(1-e^{-s_0/M^2}\right)
+{m_q\over M^2}\langle\bar{q}q\rangle\;,\label{grho}
\eeq
where we have omitted the  numerically insignificant contribution of the
gluon condensate. 

In the case of the $\rho$ off-shell, the sum rules in both structures give 
the same results, as can be seen by Eqs.~(\ref{ro1}), (\ref{ro2}), 
(\ref{phenro}) and the limit condition Eq.~(\ref{lro}). However, in the case 
of $D$ off-shell both structures give different results, as can be seen
by Eqs.~(\ref{d1}), (\ref{d2}) and (\ref{phend}). In particular,
for the structure $p_\mu$, the quark condensate also contributes with:
\beq
\Gamma_\mu^{<\bar{q}q>(D)}(p,\pli)={m_c\langle\bar{q}q\rangle\over p^2(
p^{'2}-m_c^2)}p_\mu\;,
\eeq
and the resulting sum rule in this structure is not very stable
as a function of the Borel mass. Therefore, in this work we will concentrate
in the $\pli_\mu$ structure, which we found to be the more stable one.

The parameter values used in all calculations are $m_q=(m_u+m_d)/2=7\,\MeV$, 
$m_c=1.5\,\GeV$, $m_D=1.87\,\GeV$, $m_\rho=0.77\,\GeV$, 
$\langle\overline{q}q\rangle\,=\,-(0.23)^3\,\GeV^3$. 
 We parametrize the continuum thresholds as
\beq
s_0=(m_M+\Delta_s)^2\;,\label{s0}
\eeq
where $m_M=m_D(m_\rho)$ for the case that the $\rho(D)$ meson is off-shell,
and
\beq
u_0=(m_D+\Delta_u)^2\;.\label{u0}
\eeq
The values of $u_0$ and $s_0$ are, in general, extracted from the 
two-point
function sum rules for $f_D$ and $g_\rho$ in Eqs.~(\ref{fd}) and (\ref{grho}).
Using the Borel region $3 \leq M^2\leq 5 \GeV^2$ (for the $D$  meson) and 
$0.5 \leq M^2 \leq0.9 \GeV^2$ (for the $\rho$ meson) 
we found a good stability for $f_D$ and $g_\rho$ with 
 $\Delta_s=\Delta_u\sim0.5\GeV$. In our study we will allow for a small 
variation in  $\Delta_s$ and $\Delta_u$ to test the sensitivity of our
results to the continuum contribution. 

In Fig.~1 we show the behavior of the form factor $g_{DD\rho}(Q^2)$ at 
$Q^2=1\,\GeV$ as a function of the Borel mass $\mli$, using $\Delta_s$ and 
$\Delta_u$ given in Eqs.~(\ref{s0}) and (\ref{u0}) equal to $0.5\,\GeV$. 
The dashed line gives the results for $g_{DD\rho}^{(\rho)}(Q^2)$
at a fixed ratio $\mli/M^2=m_D^2/m_D^2$, and the dot-dashed and solid lines 
give the results for $g_{DD\rho}^{(D)}(Q^2)$
at a fixed ratio $\mli/M^2=m_D^2/m_\rho^2$ (which corresponds to $M^2$
varying in the interval $0.5\leq M^2\leq0.8\GeV$). 
The dot-dashed line corresponds
to the structure  $p_\mu$  and the solid line to the structure $\pli_\mu$. As
mentioned before the sum rule in the $\pli_\mu$ structure gives a much 
flatter ``plateau'' for the form factor. 
In obtaining the results shown in Fig.~1
we have used for $f_D$ and $g_\rho$ appearing in Eqs.~(\ref{gd}) and 
(\ref{gro}) the values $f_D=200\MeV$ and $g_\rho=5.45$.
The behavior of the curve shown in Fig.~1 for other $Q^2$
and continuum threshold values is similar. 

In a recent calculation for the $D^*D\pi$ and $B^*B\pi$ form factors 
\cite{ddpi} we have included, besides the perturbative contribution, the
gluon condensate contribution. We have found out that the gluon condensate
is small, as compared with the perturbative contribution and decreases
with the Borel mass. The most important feature of the gluon condensate is 
the fact that it improves the stability of the result as a function of the 
Borel mass. Since its contribution, in the case of the $D^*D\pi$ form
factor at $M^2=5\GeV^2$, is less than 10\% of the perturbative contribution,
in this work we will neglect the gluon condensate. In order  to be sure  
that the absence of the  gluon condensate will not affect our results, 
we will extract the value of the
form factor at a higher value of the Borel mass, where we expect  the
gluon condensate contribution to be negligible. 

Fixing $\mli=4.7\,\GeV^2$ (which corresponds to $M^2=0.8\GeV^2$ for the 
case of off-shell $D$) we show, in Fig.~2,
the momentum dependence of the form factor (circles for $g_{DD\rho}^{(D)}(Q^2)$
and squares for $g_{DD\rho}^{(\rho)}(Q^2)$) in the interval $0.1\leq Q^2
\leq 5\GeV$, where we expect the sum rules to be valid (since in this 
case the cut in the $t$ channel starts at $t\sim m_c^2$ and thus the Euclidian
region stretches up to that threshold). 
In Fig.~2 we also show that the $Q^2$ dependence of the 
form factor represented by the circles  can be
well reproduced by the monopole parametrization (solid line)
\beq
g_{DD\rho}^{(D)}(Q^2)= {37.5\over Q^2+12.12}\;,
\label{mo}
\eeq
and that the form factor represented by the squares can be
well reproduced by the exponential parametrization (solid line)
\beq
g_{DD\rho}^{(\rho)}(Q^2)= 2.53e^{-{Q^2\over 0.98}}\;.
\label{exp}
\eeq

In order to extract the coupling constant 
$g_{DD\rho}$ from the form factor, we need first to define it in both cases.
In refs.~\cite{linko,osl} the coupling constant is defined as the value of the
form factor at $Q^2=0$. In this case the monopole and the exponential 
expressions should be written as:
\beq
g_{DD\rho}^{(D)}(Q^2)= g_{DD\rho}{\Lambda^2_D\over Q^2+\Lambda^2_D}\;,
\label{mo0}
\eeq
\beq
g_{DD\rho}^{(\rho)}(Q^2)= g_{DD\rho}e^{-{Q^2\over \Lambda^2_\rho}}\;.
\label{exp0}
\eeq

However, in refs.~\cite{lkz,kfs} the coupling is
defined as the value of the form factor at $Q^2=-m_M^2$, where $m_M$ is the 
mass of the off-shell meson. In this case the monopole and the exponential 
expressions should be written as:
\beq
g_{DD\rho}^{(D)}(Q^2)= g_{DD\rho}{\Lambda^2_D-m_D^2\over Q^2+\Lambda^2_D}\;,
\label{mom}
\eeq
\beq
g_{DD\rho}^{(\rho)}(Q^2)= g_{DD\rho}e^{-{Q^2+m_\rho^2\over 
\Lambda^2_\rho}}\;.
\label{expm}
\eeq

The coupling constants and cut-offs resulting of the parametrizations 
in Eqs.~(\ref{mo0}), (\ref{exp0}), (\ref{mom}) and (\ref{expm}) are given in
Table I.
\vskip 5mm
\begin{center}
\begin{tabular}{|c|c|c|c|}
\hline
&&&\\
&$g_{DD\rho}^{(M)}(Q^2=0)$ & $g_{DD\rho}^{(M)}(Q^2=-m_M^2)$ & $\Lambda_M\,
(\GeV)$\\
\hline\hline
$D$ off-shell & 3.1 & 4.4& 3.5 \\
$\rho$ off-shell & 2.5 & 4.6& 1.0 \\
\hline 
\end{tabular}
\end{center}
\begin{center}
\bf{TABLE I:} {\small Values of the coupling constants and cut-offs
which reproduce the QCDSR results for $g_{DD\rho}^{(M)}(Q^2)$.}
\end{center}
\vskip5mm

There are two very important conclusions that we can draw from the  
results in Table I. First of all is the fact that the form factor is harder 
if the 
off-shell meson is heavy, implying that the size of the vertex depends
on the exchanged meson, in agreement with our expectation, based on the 
the findings of refs.~\cite{dnn,ddpi}.
Therefore, a heavy meson will see the vertex as pointlike, whereas a 
light meson will see its extension. The second conclusion is that the value
of the coupling constant extracted from the QCDSR results depends, of course
on its definition.  Having in mind that the inherent precision
of the QCDSR method is of order of 20\%, we can say that the results nearly 
coincide for the both cases considered, i.e.  
off-shell $D$ and  $\rho$ meson. This agreement betwen the two values
is even more spectacular in the case where the coupling is extracted at
$Q^2=-m_M^2$. 

Its is also very interesting to notice that the value used in 
refs.~\cite{linko,osl}
for $g_{DD\rho}$, obtained in the framework of vector meson dominance,
is: $g_{DD\rho}=2.52$, in complete agreement with our value extracted from the
form factor normalized at $Q^2=0$. Since their form factors are also
normalized at  $Q^2=0$, we completely endorse their procedure. However,
in their analysis they vary the value of the cut-off in the range
$1\leq\Lambda_D\leq2\GeV$. Here we obtain larger values.

Using for the continuum thresholds $\Delta_s=\Delta_u=0.6\,\GeV$, in 
Eqs.~(\ref{s0}) and (\ref{u0}), the
resulting values for the coupling constants and cut-offs are given in
Table II, from where we can see that a 20\% variation in the continuum
threshods leads to less than 10\% variation in the couplings and cut-offs,
showing a good stability of the results.

\vskip 5mm
\begin{center}
\begin{tabular}{|c|c|c|c|}
\hline
&&&\\
&$g_{DD\rho}^{(M)}(Q^2=0)$ & $g_{DD\rho}^{(M)}(Q^2=-m_M^2)$ & $\Lambda_M\,
(\GeV)$\\
\hline\hline
$D$ off-shell & 3.3 & 4.4& 3.8 \\
$\rho$ off-shell & 2.6 & 4.1& 1.2 \\
\hline 
\end{tabular}
\end{center}
\begin{center}
\bf{TABLE II:} {\small Values of the coupling constants and cut-offs
which reproduce the QCDSR results for $g_{DD\rho}^{(M)}(Q^2)$
using $\Delta_s=\Delta_u=0.6\,\GeV$.}
\end{center}
\vskip5mm

Considering the incertainties in  the continuum threshold, and the difference 
in the values of the coupling constants extracted when the $D$ or $\rho$ mesons
are off-shell, our result for the $DD\rho$ coupling constant is:
\beq
g_{DD\rho}=\left\{\begin{array}{l}
            2.9\pm0.4\;\;\;\;\;\mbox{in }Q^2=0\\
            4.3\pm0.3\;\;\;\;\;\mbox{in the pole of the off-shell meson}\\
           \end{array}\right.
\eeq

\vspace{1cm}

To summarize: we have used the method of QCD sum rules  to compute form factors
and coupling constants in  $D D \rho$ vertices. Our results for the couplings 
show once more 
that this method is robust, yelding numbers which are approximately the 
same regardless of which particle we choose to be off-shell and depending 
weakly on the choice of the continuum theshold. As for the form factors,
we obtain a harder (softer) form factor when the off-shell particle is 
heavier (lighter).  This confirms our expectation based on previous works 
and answers the question in the title: heavy mesons ``see'' smaller sizes 
in the light mesons, but the latter can not resolve small structures in 
the former. Our program will continue and studies of other vertices are  
in progress.

\vspace{1cm}
 
\underline{Acknowledgements}: 
This work has been supported by CNPq and FAPESP.

\begin{figure} \label{fig1}
\begin{center}
\epsfysize=9.0cm
\epsffile{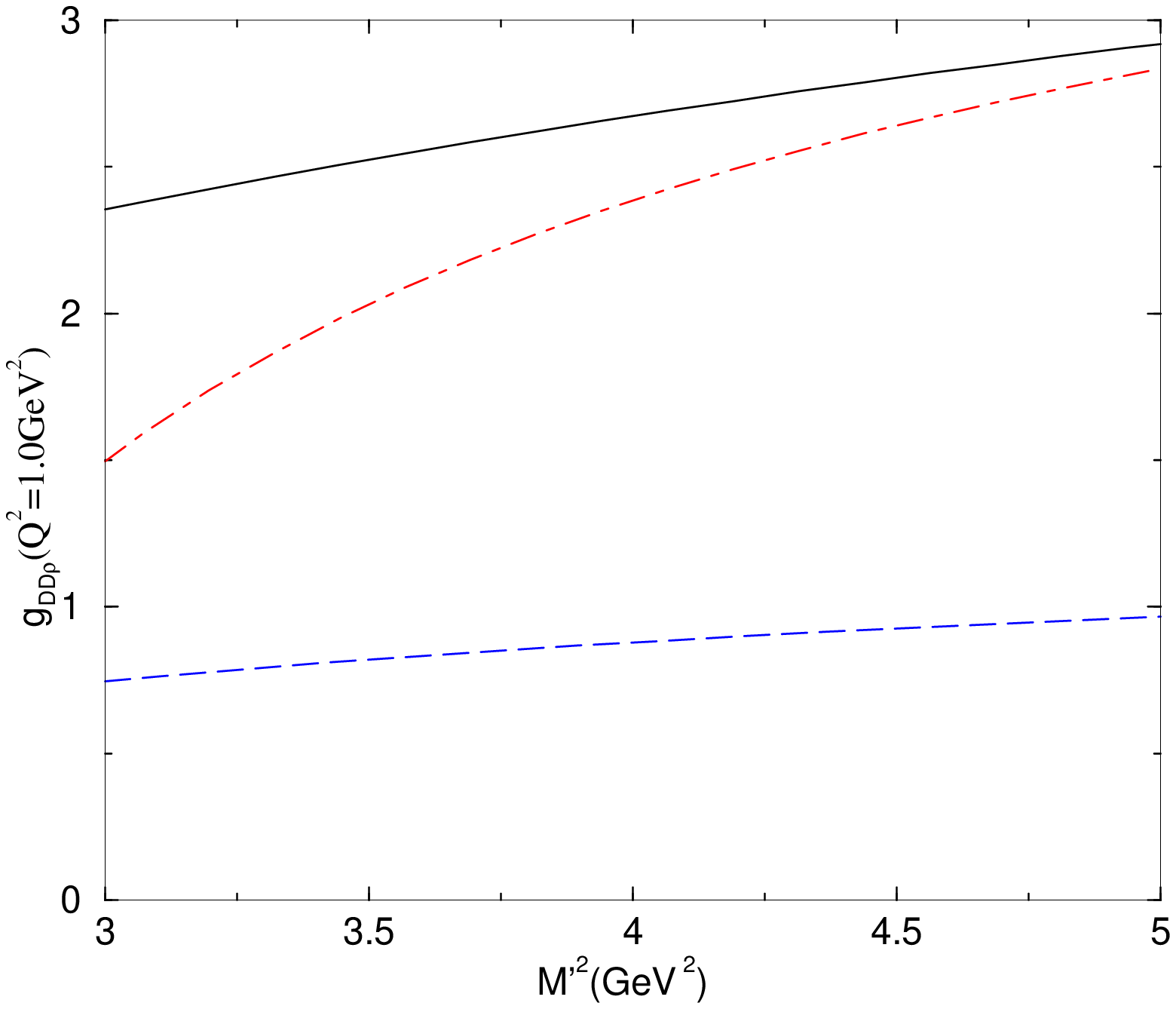}
\caption{$\mli$ dependence of the $DD\rho$ form factors at $Q^2=1\,\GeV^2$ 
for $\Delta_s=\Delta_u=0.5\,\GeV$. The dashed line gives the QCDSR result
for $g_{DD\rho}^{(\rho)}(Q^2)$ and the dot-dashed and solid lines give
the QCDSR results for $g_{DD\rho}^{(D)}(Q^2)$ in the $p_\mu$ and $\pli_\mu$
structures respectively.}
\end{center}
\end{figure}

\begin{figure} \label{fig2}
\begin{center}
\vskip -1cm
\epsfysize=9.0cm
\epsffile{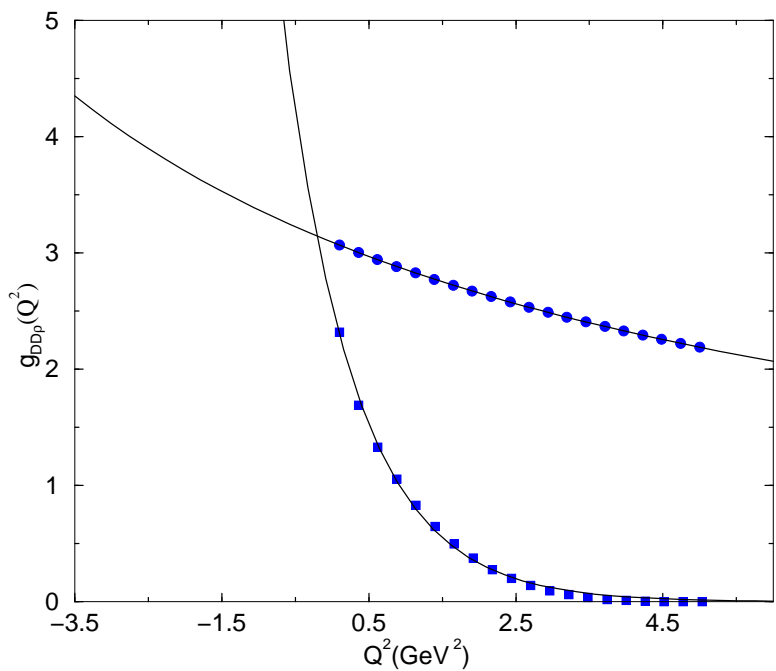}
\caption{Momentum dependence of the $DD\rho$ form factor for
$\Delta_s=\Delta_u=0.5\,\GeV$. The solid lines give the 
parametrization of the QCDSR results through Eq.~(\protect\ref{mo}) for the 
circles, and Eq.~(\protect\ref{exp}) for the squares.}
\end{center}
\end{figure}


\vspace{0.5cm}


\begin{thebibliography}{99}

\bibitem{bad} For a very good introduction,  see R.K. Bhaduri, ``Models of
the Nucleon'', Lecture Notes and Supplements in Physics, Addison-Wesley 
Pub. Co., 1988.

\bibitem{pauli} For recent papers addressing the subject, see for example,
                H.C. Pauli, hep-ph/0107302;  H.C. Pauli and A. Mukherjee, 
                hep-ph/0103150.

\bibitem{cart} See for example, N. Cartiglia, hep-ph/9703245, talk given 
               at the SLAC Summer School 1996.

\bibitem{linko} Z. Lin and C.M. Ko, 
                 {\sl Phys. Rev.} {\bf C62}, 034903 (2000). 

\bibitem{lkz} Z. Lin, C.M. Ko and B. Zhang,
              {\sl Phys. Rev.} {\bf C61}, 024904 (2000).

\bibitem{ldk}  ``Charm meson scattering cross sections by pion and rho 
                meson'', nucl-th/0006086,  Z. Lin, T. G. Di and C. M. Ko.

\bibitem{haglin} Kevin L. Haglin, {\sl Phys. Rev.} {\bf C61}, 
                 031902 (2000); Kevin L. Haglin and Charles Gale, {\sl Phys. Rev}
                 {\bf C61} 065201 (2001).

\bibitem{haga} ``Hadronic interactions of the $ J/\psi$'', nucl-th/0010017, 
               Kevin L. Haglin and C. Gale.


\bibitem{stt} A. Sibirtsev, K. Tsushima and A. W. Thomas,  
               {\sl Phys. Rev.} {\bf  C63}  044906 (2001). 


\bibitem{osl} Y. Oh, T. Song and S.H. Lee, {\sl Phys. Rev.}
              {\bf C63}, 034901 (2001).


\bibitem{nnr} F.S. Navarra, M. Nielsen and M.R. Robilotta,   
              {\sl Phys. Rev.} {\bf C 64}, 021901 R (2001).

\bibitem{lkl} ``Charmonium  absorption cross section by nucleon'', 
              nucl-th/0107058, W.Liu, C.M. Ko and Z.W. Lin.


\bibitem{nn}   F.S. Navarra and M. Nielsen, 
               {\sl Phys. Lett.} {\bf B443}, 285 (1998).

\bibitem{dnn} F.O. Dur\~aes, F.S. Navarra and M. Nielsen, {\sl Phys. Lett.}  
              {\bf B498},  169  (2001). 


\bibitem{ddpi} F.S. Navarra, M. Nielsen, M.E. Bracco, M. Chiapparini and
               C.L. Schat, {\sl Phys. Lett.}  {\bf B489},  319  (2000). 


\bibitem{io2}  B.L. Ioffe and A.V. Smilga, {\sl Nucl. Phys.} {\bf B216} 373
               (1983); {\sl Phys. Lett.} {\bf B114}, 353 (1982).


\bibitem{kfs} W. Koepf, L.L. Frankfurt and M. Strikman, {\sl Phys. Rev.}
              {\bf D53}, 2586 (1996).



\end{thebibliography}
\end{document}